\documentclass[doublecol]{epl2}

\usepackage{graphicx}
\usepackage{subfigure}
\usepackage{amsthm}
\usepackage{amsmath}
\usepackage{amssymb}
\usepackage{verbatim}
\usepackage{dcolumn}% Align table columns on decimal point
\usepackage{bm}% bold math
\usepackage{epsf}
\usepackage{color}
\usepackage[colorlinks=true,citecolor=blue,linkcolor=blue,urlcolor=blue]{hyperref}%
\usepackage{xcolor}
\usepackage{dsfont}

\newcommand{\pd}{\partial}

\newcommand{\e}[1]{\exp{\left(#1\right)}}
\newcommand{\lo}[1]{\ln{\left(#1\right)}}

\newcommand{\bla}{bla\\bla\\bla\\bla\\bla}

\newcommand{\mc}[1]{\mathcal{#1}}

\newcommand{\mrm}[1]{\mathrm{#1}}

\title{Boosting thermodynamic performance by bending space-time}

\author{Emily E. Ferketic \and Sebastian Deffner \thanks{E-mail: \email{deffner@umbc.edu}}}
\shortauthor{Emily E. Ferketic and Sebastian Deffner}
\institute{Department of Physics, University of Maryland, Baltimore County, Baltimore, MD 21250, USA}

\abstract{Black holes are arguably the most extreme regions of the universe. Yet, they are also utterly inaccessible to experimentation, and even just indirect observation poses significant technical challenges. The phenomenological approach of thermodynamics is uniquely suited to explore at least some of the physical properties of such scenarios, and this has motivated the study of so-called holographic engines. We show that the efficiency of an endoreversible Brayton cycle is given by the Curzon-Ahlborn efficiency if the engine is fueled by a 2-dimensional ideal gas; and that the efficiency is higher, if the working medium is a (2+1)-dimensional BTZ black hole. These findings may be relevant not only in the quest to unlock the mysteries of black holes, but also for potential technological applications of graphene.
}

\begin{document}

\maketitle

As pretentious as it might sound,  every single time we navigate our cars towards a gas station we are forced to make a decision of non-trivial thermodynamic consequences -- which fuel should we buy? Typically, gas stations offer the mundane garden variety of rather similar chemical compositions, which offer only small differences in maximal power and engine efficiency.

The situation is a lot more interesting at the forefront of technological development, where more and more quantum fuels are being investigated \cite{Myers2020AVS}.  In fact, it has been recognized early on that engines operating with quantum fuels can do things that classical engines cannot \cite{Scully2002,Scully2003},  and that this is perfectly legal under the laws of thermodynamics \cite{Deffner2019book}.  As a phenomenological theory, thermodynamics only has very limited knowledge of the underlying microscopic structure of the considered systems \cite{Callen1985}. Any differences in engine performance are thus fully derived from the fundamental relation,  or a complete set of equations of state. This fundamental relation can be obtained empirically as, for instance, for the ideal gas, or it can be derived from microscopic theories such as statistical mechanics or quantum field theory \cite{Banados1992PRL}.

Owing to the universality of the thermodynamic laws, deriving statements about the thermodynamic characteristics becomes then almost only an exercise. However, such an approach does permit to unveil further statements about, e.g., the question which fuel has the best performance in what scenario. In particular, we have recently shown that endoreversible Otto engines can outperform classical engines, if they operate with quantum fuels \cite{Deffner2018Entropy,Smith2020JNET}, whereas corrections from special relativity do not support an additional boost of thermodynamic performance \cite{Myers2021NJP}.

Thus, the rather natural question arises whether in even more extreme scenarios thermodynamic performance is governed by the properties of space-time. Such a question is not quite as esoteric as it might appear, since black hole thermodynamics \cite{Bekenstein1980PT,Carlip2014IJMP} has been a particular fruitful approach in elucidating the properties of such mysterious regions of space in the universe. The central quantity is the Bekenstein-Hawking entropy \cite{Hawking1971,Bekenstein1972,Bekenstein1973}, which is the amount of entropy that must be assigned to a black hole in order for it to comply with the laws of thermodynamics as they are interpreted by external observers. To further investigate the thermodynamic properties of such black holes, it has now become an almost common practice to analyze so-called \emph{holographic engines} \cite{Johnson2014CQG}. These are engine cycles, in which the cosmological constant, i.e, the energy density of the considered region of space \cite{Carrol1992ARAA}, is treated as thermodynamic variable akin to pressure. Such research is often motivated by attempts to develop a more complete understanding of the universe \cite{Johnson2014CQG}. Yet, it is somewhat unlikely that our civilization will ever reach a state in which one would use cosmological black holes as fuel in technological applications \cite{Opatrny2012AJP}. However, special and general relativity have found close analogies in condensed matter \cite{Wehling2014AP} and optical systems \cite{Philbin2008Science,Braidotti2022AVS}, which are clearly going to find near-term applications in nanotechnology. 

In the present analysis, we thus investigate whether singularities in space-time can be exploited as a thermodynamic resource. More specifically, we show that the efficiency at maximum power of an endoreversible Brayton cycle is higher, when operated with a Ba\~{n}ados-Teitelboim-Zanelli (BTZ) black hole \cite{Banados1992PRL} than when fueled by an ideal gas. The Brayton cycle consists of two isentropic and two isobaric strokes, and hence it is most conveniently treated in the entalphy representation \cite{Callen1985}. Since also the fundamental relation of black holes is particularly accessible in enthalpy form \cite{Kastor2009CQG,Kastor2018CQG}, the Brayton cycle appears as the somewhat natural choice for holographic engines.  As main results, we find that for the 2-dimensional ideal gas the efficiency at maximum power is identical to the Curzon-Ahlborn efficiency \cite{Curzon1975}, and that for the BTZ black hole the efficiency is larger. All results are derived analytically and in closed form.

\section{Endoreversible Brayton cycle}

We start by establishing notions and notations. The ideal \emph{Brayton or Joule cycle} is consists of two isentropic and two isobaric strokes \cite{Callen1985}. It is closely related to the Otto cycle, in that the two isochoric strokes are replaced by two isobaric strokes, which is more apt as a design principle for, e.g., gas turbines.  The endoreversible Brayton cycle can then be constructed in full analogy to the endoreversible Otto cycle \cite{Deffner2018Entropy}.  In particular, we have:

\subsection{Isobaric heating (A$\rightarrow$ B)}

For the isobaric strokes it is most convenient to work with the enthalpy representation of the fundamental equation, $H=H(S,P)$. Recall that in differential form we have \cite{Callen1985},
\begin{equation}
dH=T\, dS+V \,dP\quad \text{and\, thus}\quad Q_{A\rightarrow B}=H_B-H_A\,.
\end{equation}
Correspondingly, we can express the work done during isobaric heating as
\begin{equation}
W_{A\rightarrow B}=\Delta E_{A\rightarrow B} -Q_{A\rightarrow B}= P_A\left(V_A-V_B\right)
\end{equation}
which is nothing but the first law of thermodynamics, and where we used the thermodynamic identity $H=E+PV$, and that $P_A=P_B$.

In complete analogy to the endoreversible Carnot \cite{Curzon1975} and Otto \cite{Deffner2018Entropy,Smith2020JNET,Myers2021NJP} cycles, we now assume that the working substance is in a state of \emph{local} equilibrium at a temperature below the temperature of the hot reservoir.  Thus we write,
\begin{equation}
T(0)=T_A\quad\mrm{and}\quad T(\tau)=T_B\quad\mrm{with}\quad T_A<T_B \leq T_\mrm{hot}\,,
\end{equation}
where $\tau$ is the finite duration of the stroke.  In linear approximation,  the time-dependence of the temperature is then given by Fourier's law \cite{Deffner2018Entropy},
\begin{equation}
\label{eq:fourier_hot}
\frac{d T}{dt}=-\alpha \left(T(t)-T_\mrm{hot}\right)
\end{equation}
where $\alpha$ is a constant depending on the heat conductivity and heat capacity of the working substance. 

Equation~\eqref{eq:fourier_hot} can be solved exactly, and we obtain
\begin{equation}
\label{eq:rel_hot}
T_B-T_\mrm{hot}=\left(T_A-T_\mrm{hot}\right)\,\e{-\alpha\, \tau}\,.
\end{equation}

\subsection{Isentropic expansion (B $\rightarrow C)$}

During the isentropic strokes the working substance is decoupled from the environment. Therefore,  the endoreversible description is identical to the equilibrium cycle. Again from the first law of thermodynamics, $\Delta E=Q+W$, we simply have,
\begin{equation}
Q_{B\rightarrow C}=0\quad\mrm{and}\quad W_{B\rightarrow C}=E_C-E_B
\end{equation}
where $Q_{B\rightarrow C}$ is the heat exchanged, and $W_{B\rightarrow C}$ is the work performed during the expansion.

\subsection{Isobaric cooling (C $\rightarrow$ D)}

Heat and work during the isobaric cooling read,
\begin{equation}
Q_{C\rightarrow D}=H_D-H_C\quad\mrm{and}\quad W_{C\rightarrow D}= P_C\left(V_C-V_D\right)\,,
\end{equation}
and we now have
\begin{equation}
T(0)=T_C\quad\mrm{and}\quad T(\tau)=T_D\quad\mrm{with}\quad T_C>T_D \geq T_\mrm{cold}\,.
\end{equation}
For the sake of simplicity,  we assume that the stroke time of the isobaric cooling is identical to the stroke time of the isobaric heating, $\tau$.  The generalization to different strokes times is straight forward, but does significantly increase the clutter in the formulas, cf Ref.~\cite{Deffner2018Entropy,Smith2020JNET} for the corresponding Otto cycle.

Similarly to above \eqref{eq:fourier_hot} the heat transfer is described by Fourier's law
\begin{equation}
\label{eq:fourier_cold} 
\frac{d T}{dt}=-\alpha  \left(T(t)-T_\mrm{cold}\right)\,,
\end{equation}
where $\alpha$ is a constant characteristic for the cold stroke., which again we choose to be identical to the heating stroke.  From the solution of Eq.~\eqref{eq:fourier_cold} we now obtain
\begin{equation}
\label{eq:rel_cold}
T_D-T_\mrm{cold}=\left(T_C-T_\mrm{cold}\right)\,\e{-\alpha\, \tau }\,,
\end{equation}
which describes the cooling from $T_C$ to $T_D$.

\subsection{Isentropic compression (D $\rightarrow$ A)}

The cycle is completed by isentropic  expansion, for which we have
\begin{equation}
Q_{D\rightarrow A}=0\ \quad\mrm{and}\quad W_{D\rightarrow A}=E_D-E_A\,.
\end{equation}

\subsection{Brayton efficiency} As always, the efficiency of a thermodynamic cycle is defined as ratio of work output, $W_\mrm{out}=(Q_{A\rightarrow B}-Q_{C\rightarrow D})$,  and (hot) heat input,  $Q_{A\rightarrow B}$.  For the present Brayton cycle we immediately find
\begin{equation}
\eta=1+\frac{Q_{C\rightarrow D}}{Q_{A\rightarrow B}}=1-\frac{H_C-H_D}{H_B-H_A}\,,
\end{equation}
which cannot be further simplified without knowledge of the fundamental equation, $H=H(S,P)$.

\section{The ideal gas}

\begin{figure}
\includegraphics[width=.48\textwidth]{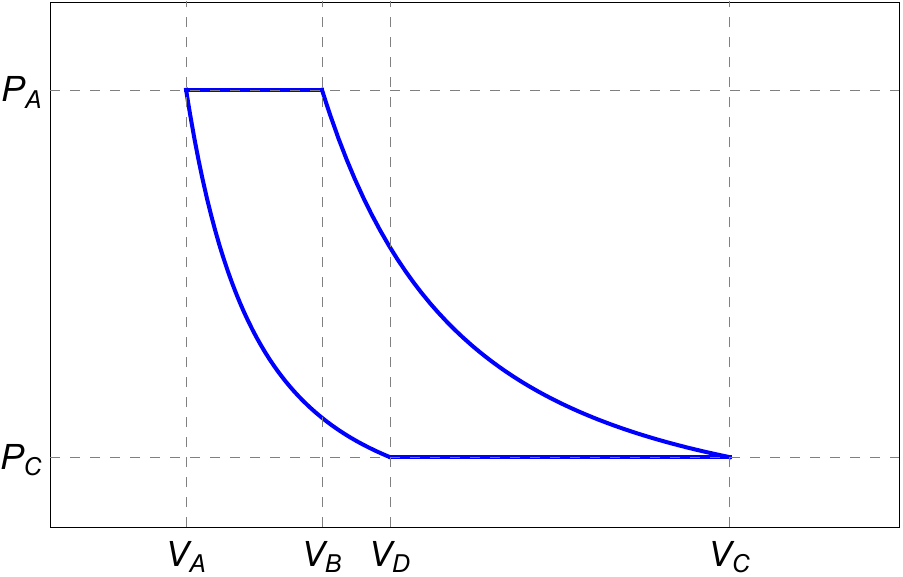}
\includegraphics[width=.48\textwidth]{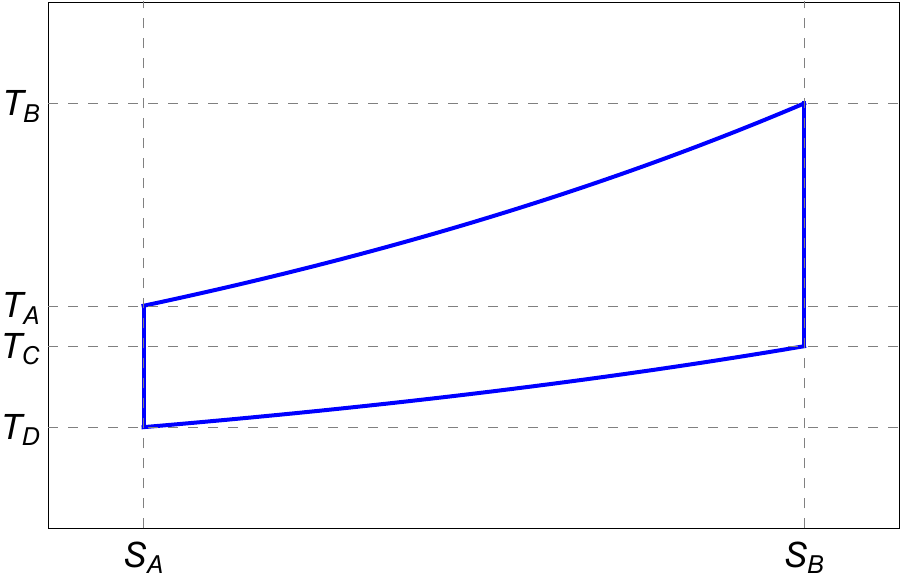}
\caption{\label{fig:ideal} Pressure-volume (top) and temperature-entropy (bottom) diagrams of the Brayton cycle for an ideal gas. }
\end{figure}

As a point of reference and to build intuition we begin with an endoreversible Brayton cycle that operates with an ideal gas as working medium.  The well-known equation of state,  aka the ideal gas law reads
\begin{equation}
\label{eq:ideal_gas}
P V= N k_B T\,,
\end{equation}
where $k_B$ is Boltzmann's constant. Expressing the entropy as a function of temperature \cite{Callen1985}
\begin{equation}
\label{eq:ent_ideal}
S=S_0+c N k_B \lo{\frac{T}{T_0}}+N k_B \lo{\frac{V}{V_0}}
\end{equation}
we obtain the corresponding PV and TS-diagrams depicted in Fig.~\ref{fig:ideal}. Note that $S_0$,  $T_0$, $V_0$, and  $c$ are constants that only depend on the dimension of the system. Below we will be considering the BTZ black hole, which lives in $(2+1)$-dimensional space. For a fair comparison, we continue in only 2 spatial dimensions, and hence $c=1$ \cite{Callen1985}.

From the expression of the entropy \eqref{eq:eff_ideal} and the ideal gas law \eqref{eq:ideal_gas}, we can also directly read off that
\begin{equation}
\label{eq:com}
\frac{T_C}{T_B}=\frac{T_D}{T_A}=\frac{P_C}{P_A}\equiv\kappa
\end{equation}
where $\kappa$ is the compressor ratio.

It is also convenient to express the enthalpy as a function of temperature. We have
\begin{equation}
\label{eq:enth_ideal}
H=E+PV=2 N k_B T=c_P T\,,
\end{equation}
where we introduced the isobaric heat capacity $c_P\equiv 2 N k_B$. Hence we obtain for the efficiency
\begin{equation}
\label{eq:eff_ideal}
\eta=1-\frac{T_C-T_D}{T_B-T_A}=1-\kappa\,.
\end{equation}
where we used Eq.~\eqref{eq:com}. Thus, we are now interested in determining the value of $\kappa$,  for which the power output of the endoreversible engine is maximum.

The power output is given by the work output divided by the total cycle time $\tau_\mrm{cyc}=\gamma \tau$,
\begin{equation}
\mc{P}=\frac{W_\mrm{out}}{\tau_\mrm{cyc}}=\frac{c_P}{\gamma \tau}\left(T_B-T_A+T_D-T_C\right)\,.
\end{equation}
where we used again $W_\mrm{out}=(Q_{A\rightarrow B}-Q_{C\rightarrow D})$ and Eq.~\eqref{eq:com}. Now further exploiting Eqs.~\eqref{eq:fourier_hot} and \eqref{eq:fourier_cold} it is a little tedious, but straight forward to eliminate the temperatures $T_A$, $T_B$, $T_C$, and $T_D$. We obtain
\begin{equation}
\label{eq:pow_ideal}
\mc{P}=\frac{c_P\tanh\left(\alpha\tau/2\right)}{\gamma \tau}\,\frac{\kappa-1}{\kappa}\left(T_\mrm{hot} \kappa -T_\mrm{cold}\right)\,.
\end{equation}
Notice that the expression for the power \eqref{eq:pow_ideal} factorizes, and thus $\mc{P}$ can be maximized as a function of $\kappa$ for all values of $\alpha$ and $\tau$.  This expression for the power \eqref{eq:pow_ideal} has a rather simple dependence on $\kappa$ with a unique maximum.
\begin{figure}
\includegraphics[width=.48\textwidth]{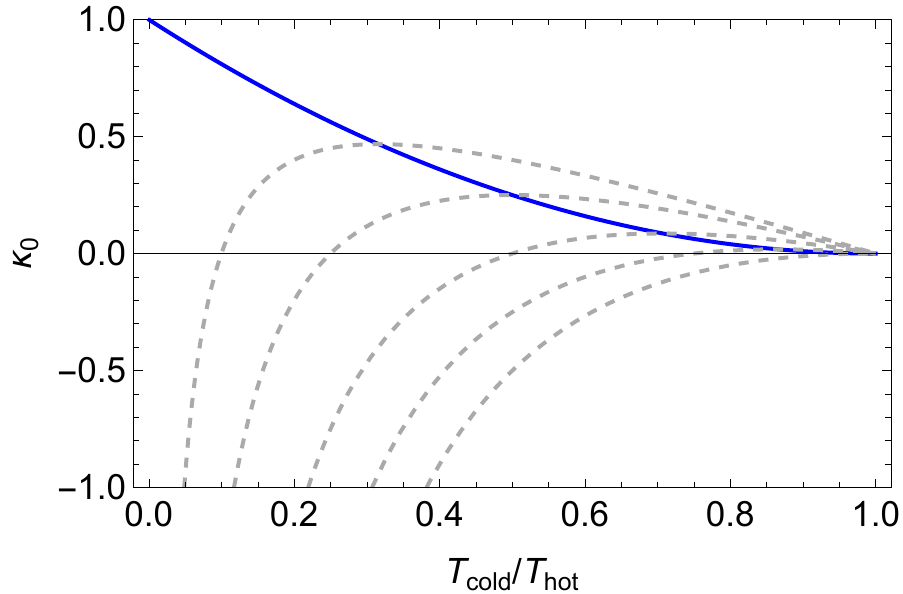}
\caption{\label{fig:kappa_ideal} Compressor ratio $\kappa_0$, for which $\mc{P}(\kappa_0)$ is maximum as a function of $T_\mrm{cold}/T_\mrm{hot}$ (blue, solid line), cf. Eq.~\eqref{eq:pow_max}. The gray dashed lines correspond to  $\mc{P}(\kappa)$ \eqref{eq:pow_ideal} in units of $c_P\tanh\left(\alpha\tau/2\right)/\gamma \tau$ for different values of $T_\mrm{hot}$ and serve to demonstrate the well-defined maximum.}
\end{figure}
As usual, the maximum of $\mc{P}(\kappa)$ is determined by
\begin{equation}
\label{eq:pow_max}
\frac{\pd \mc{P}}{\pd \kappa}=0\quad\text{if\, and\, only\, if}\quad \kappa=\kappa_0=\sqrt{\frac{T_\mrm{cold}}{T_\mrm{hot}}}\,.
\end{equation}
This is illustrated in Fig.~\ref{fig:kappa_ideal}.

In conclusion, we obtain that the efficiency of an endoreversible Brayton cycle at maximal power is given for a 2-dimensional ideal gas by
\begin{equation}
\label{eq:curz}
\eta=1-\sqrt{\frac{T_\mrm{cold}}{T_\mrm{hot}}}
\end{equation}
which is nothing but the Curzon-Ahlborn efficiency \cite{Curzon1975}. This result is remarkable, since in the original treatment of Curzon and Ahlborn \cite{Curzon1975} the analysis was restricted to endoreversible Carnot cycles.

However, we have also already shown that the endoreversible Otto engines have the same efficiency at maximal power, as long as the internal energy is a linear function of the temperature \cite{Smith2020JNET}.  In the present case, also the enthalpy is linear in temperature and thus Eq.~\eqref{eq:curz} is mathematically not all that surprising.  In fact, the ideal gas is rather special, since generally the enthalpy is not simply proportional to the internal energy, and we would thus expect the efficiency to intimately depend on the corresponding equations of state. 

However, Eq.~\eqref{eq:curz} does pose the interesting question what properties of working mediums can be exploited to enhance thermodynamic efficiency. In the following, we will demonstrated that ``bending space-time'' is such a ``thermodynamic resource''.

\section{The BTZ black hole}

Rather than continuing the analysis for a realistic model with curved space-time, we analyze a pedagogical example of some physical significance. In fact, the BTZ black hole, named after Ba\~{n}ados, Teitelboim, and Zanelli \cite{Banados1992PRL}, has proven to be particularly instructive \cite{Carlip1995CQG,Birmingham2001IJMP,Dolan2011CQG,Singh2014JPCS} due to its mathematical simplicity. This ``black hole'' is a solution for (2+1)-dimensional topological gravity with a negative cosmological constant \cite{Banados1992PRL}. Remarkably, many of the physical properties of the BTZ black hole mirror the properties of more involved models, such as the (3+1)-dimensional Schwarzschild and Kerr black holes \cite{Carlip1995CQG}. In particular, it supports a version of the \emph{no hair theorem} \cite{Birmingham2001IJMP} and it has a consistent thermodynamic description \cite{Dolan2011CQG}.

As usual in black hole thermodynamics \cite{Carlip2014IJMP} one identifies the mass with the thermodynamic entropy, and the cosmological constant plays the role of pressure \cite{Dolan2011CQG}. Note that the cosmological constant is not a ``constant'' in the strict sense, but rather represents the energy density of the considered space \cite{Carrol1992ARAA}. Hence, this black hole pressure can be considered a thermodynamic variable, that can be varied in black hole heat engines \cite{Opatrny2012AJP,Johnson2016CQG_2,Johnson2018CQG_2,Bhamidipati2017EPJC,Hendi2018PLB,Chakraborty2018IJMP,
Chakraborty2018IJMP_2,Wei2019NPB,Yerra2019MPLA,Guo2021MPLA,Feng2021CTP} and so-called holographic engines \cite{Johnson2014CQG,Johnson2016entropy,Johnson2016CQG,Hennigar2017CQG,Johnson2018CQG,Zhang2018EPJC,Mo2018JHEP,
Zhang2019JHEP,Ahmed2019CQG,Rosso2019IJMP,Ghaffarnejad2020NPB,Panah2020NPB,Sun2021CPB,Moumni2021NPB}.

We will now explore whether the efficiency of an endoreversible Brayton cycle with such a BTZ black hole as a working medium has a higher efficiency than we found above for the ideal gas. The BTZ enthalpy is simply given by \cite{Dolan2011CQG}
\begin{equation}
\label{eq:enth_BTZ}
H(S,P)=\frac{4 \ell^2}{\pi} S^2 P\,,
\end{equation}
where $\ell\equiv \hbar G_N$ with the gravitational constant $G_N$.

\begin{figure}
\includegraphics[width=.48\textwidth]{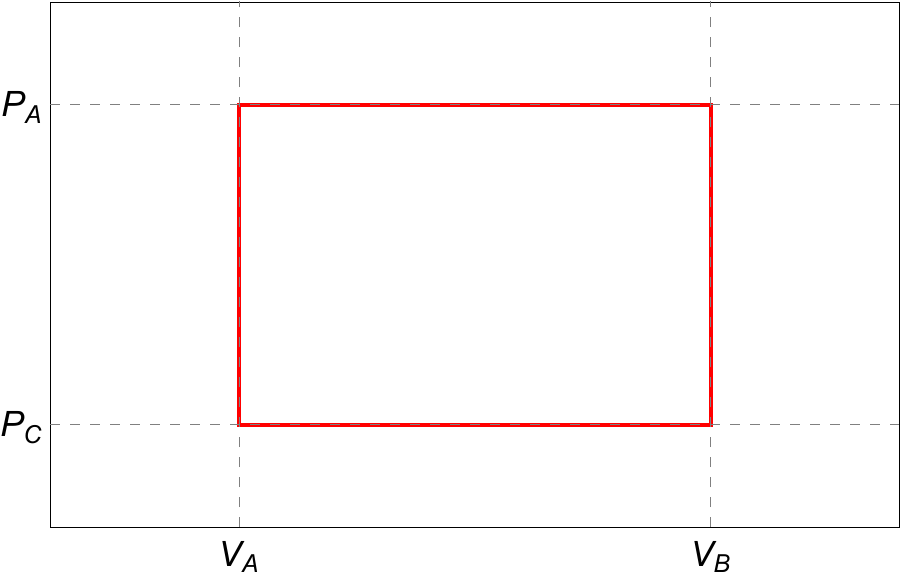}
\includegraphics[width=.48\textwidth]{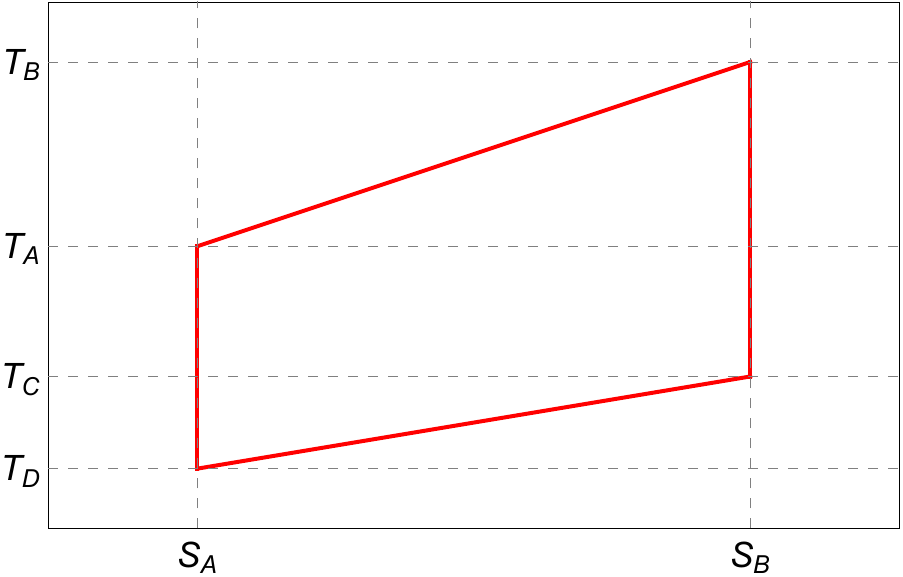}
\caption{\label{fig:BTZ} Pressure-volume (top) and temperature-entropy (bottom) diagrams of the Brayton cycle for a BTZ black hole. }
\end{figure}

As before, we first determine the $PV$- and $TS$-diagrams for the corresponding Barton cycle. Using thermodynamic identities \cite{Callen1985} we have for the volume,
\begin{equation}
V(S,P)=\frac{\pd H}{\pd P}\bigg|_S=\frac{4 \ell^2}{\pi} S^2\,.
\end{equation}
Hence, the $PV$-diagram becomes a simple rectangle, cf. Fig.~\ref{fig:BTZ} in the top panel.

Now exploiting the Euler relation \cite{Callen1985}
\begin{equation}
E=TS-PV\quad\text{and}\quad H=TS
\end{equation}
we obtain the mechanical equation of state
\begin{equation}
P \sqrt{V}=\frac{2\ell}{\sqrt{\pi}} T\,,
\end{equation}
which replaces the ideal gas law \eqref{eq:ideal_gas}. The latter also permits to express the entropy as a function of temperature, $T$, and pressure, $P$. We have
\begin{equation}
\label{eq:ent_BTZ}
S=\frac{\pi}{4 \ell^2}\frac{T}{P}\,,
\end{equation}
and thus isobaric processes are straight lines in the $TS$-diagram, cf. Fig.~\ref{fig:BTZ} in the bottom panel. Thus, while deriving the enthalpy from general relativity is a somewhat involved task \cite{Dolan2011CQG}, the resulting thermodynamic cycle could hardly be any simpler.

From Eq.~\eqref{eq:ent_BTZ} it is also easy to see that we again have
\begin{equation}
\label{eq:com_2}
\frac{T_C}{T_B}=\frac{T_D}{T_A}=\frac{P_C}{P_A}\equiv\kappa\,.
\end{equation}
Accordingly, the total work $W_\mrm{out}=(Q_{A\rightarrow B}-Q_{C\rightarrow D})$ can be written as
\begin{equation}
W_\mrm{out}=\frac{\pi}{4 \ell^2} \frac{1-\kappa}{P_A} \left(T_A^2-T_B^2\right)\,.
\end{equation}
Similarly, we find
\begin{equation}
Q_{A\rightarrow B}=H_B-H_A=\frac{\pi}{4 \ell^2} \frac{1}{P_A} \left(T_A^2-T_B^2\right)\,,
\end{equation}
and hence the efficiency can again be written as
\begin{equation}
\eta=\frac{W_\mrm{out}}{Q_{A\rightarrow B}}=1-\kappa\,.
\end{equation}
The task is now to determine the value of $\kappa$ for which the power output is maximum.

As before, we now exploit Fourier's law \eqref{eq:fourier_hot} and \eqref{eq:fourier_cold} to eliminate $T_A$ and $T_B$. After straight forward manipulations we obtain
\begin{equation}
\label{eq:pow_BTZ}
\mc{P}=\frac{\pi \,\tanh\left(\alpha\tau/2\right)}{8\ell^2\,\gamma\,\tau\, P_A}\,  \frac{\kappa-1}{\kappa^2} \left(T_\mrm{hot}^2\kappa^2-T_\mrm{cold}^2\right)\,,
\end{equation}
which is similar to what we found for the ideal gas \eqref{eq:pow_ideal}.

\begin{figure}
\includegraphics[width=.48\textwidth]{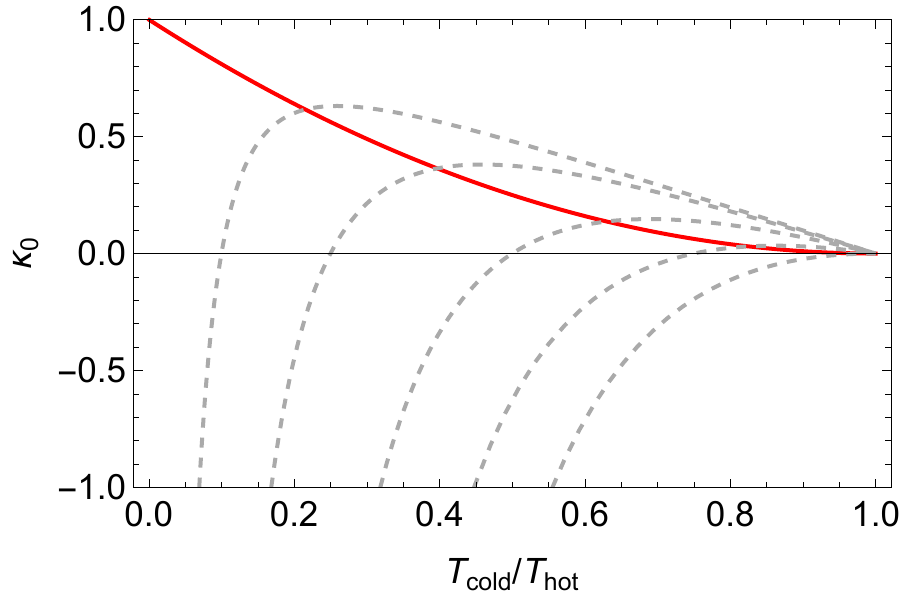}
\caption{\label{fig:kappa_BTZ}Compressor ratio $\kappa_0$, for which $\mc{P}(\kappa_0)$ is maximum as a function of $T_\mrm{cold}/T_\mrm{hot}$ (blue, solid line), cf. Eq.~\eqref{eq:kappa}. The gray dashed lines correspond to  $\mc{P}(\kappa)$ \eqref{eq:pow_BTZ} in units of $\pi \,\tanh\left(\alpha\tau/2\right)/8\ell^2\,\gamma\,\tau\, P_A$ for different values of $T_\mrm{hot}$ and serve to demonstrate the well-defined maximum.}
\end{figure}

Remarkably, the expression for the power \eqref{eq:pow_BTZ} also factorizes, and hence $\mc{P}$ can be maximized as a function of only the compressor ratio $\kappa$. However, whereas for the ideal gas the resulting polynomial was only quadratic \eqref{eq:pow_max}, for the present case the situation is a little more involved. In fact, the power has extrema for all $\kappa$ that solve
\begin{equation}
\kappa^3\,T_\mrm{hot}^2+(\kappa-2)\, T_\mrm{cold}^2=0\,.
\end{equation}
This third order polynomial can be solved analytically, but only one solution is real (the other two are complex).  For this unique, real solution the power \eqref{eq:pow_BTZ} as a well-defined maximum, cf. Fig.~\eqref{fig:kappa_BTZ}.
This maximum is attained for $\kappa=\kappa_0$, where
\begin{equation}
\label{eq:kappa}
\kappa_0=\frac{-3^{1/3}\,T^{2/3}_\mrm{cold}+\left(9\, T_\mrm{hot}+\sqrt{3}\sqrt{T_\mrm{cold}^2+27\, T_\mrm{hot}^2}\right)^{2/3}}{3^{2/3}\,T_\mrm{hot} T_\mrm{cold}^{-2/3} \left(9\, T_\mrm{hot}+\sqrt{3}\sqrt{T_\mrm{cold}^2+27\, T_\mrm{hot}^2}\right)^{1/3} }\,.
\end{equation}
However, the resulting expression is involved enough that it is not immediately clear how it compares to our findings for the ideal gas \eqref{eq:pow_max}.

\begin{figure}
\includegraphics[width=.48\textwidth]{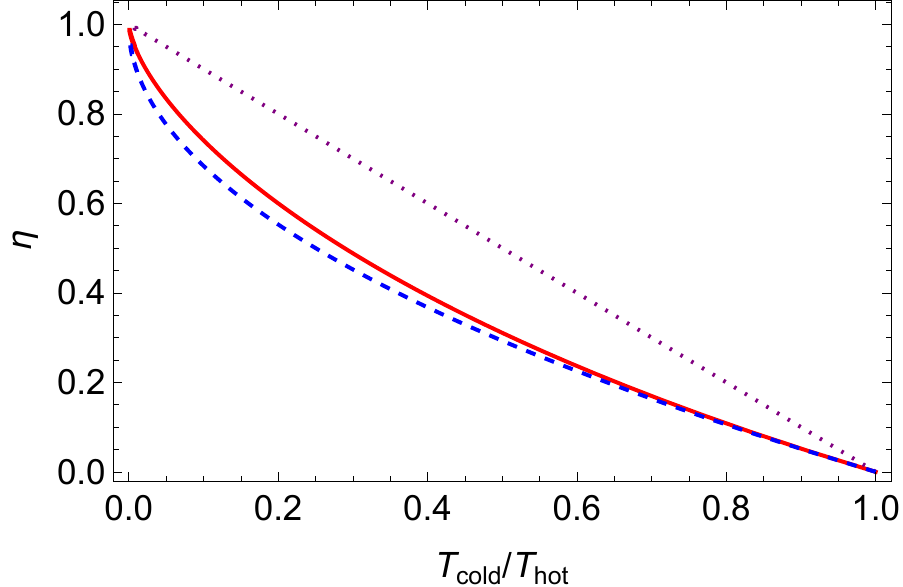}
\caption{\label{fig:eff} Efficiency of the endoreversible Brayton cycle at maximum power for the ideal gas \eqref{eq:curz} (blue dashed, line), and the BTZ black hole \eqref{eq:eff_BTZ} (red, solid line). The dashed, purple line depicts the Carnot efficiency, $\eta_\mrm{CA}=1-T_\mrm{cold}/T_\mrm{hot}$, for comparison. }
\end{figure}

The corresponding efficiency at maximum power is given by
\begin{equation}
\label{eq:eff_BTZ}
\eta=1-\kappa_0\,.
\end{equation}
This is plotted together with the efficiency for the ideal gas \eqref{eq:curz} in Fig.~\eqref{fig:eff}.
We observe that the efficiency for the BTZ black hole \eqref{eq:eff_BTZ} is larger than the Curzon-Ahlborn efficiency, which we found for the endoreversible Brayton cycle operating with an ideal gas. As a consistency check, we also compare our results with Carnot efficiency, $\eta_\mrm{CA}$. The endorervesible results are clearly below $\eta_\mrm{CA}$ as they should. In conclusion, we have found that the endoreversible Brayton cycle is indeed more efficient at maximum power, if a BTZ black hole is used as working medium. Since the BTZ black hole shares all important characteristics of more realistic black holes, we may also conclude that singularities in space-time can, indeed, enhance thermodynamic performance. 

\section{Concluding remarks}

Thermodynamics was invented at the beginning of the Industrial Revolution in an attempt to understand and optimize newly emerging technologies -- steam engines. Therefore, most of the thermodynamic language is phrased for devices that operate with gases, and observables such as pressure are the central thermodynamic variables. However, the mindset and approach of thermodynamics are as universal as they could possibly be. As a phenomenological theory, it is often the only approach that permits to investigate physical systems that are not accessible to experimentation. This is particularly true for black holes, with the caveat that then the interpretation of the results may become somewhat involved. For instance, one may want to ponder what identifying a negative cosmological constant with a variable pressure really means.

However, it also has become more and more apparent that many inaccessible objects in the universe have close analogies in condensed matter, and in particular in 2-dimensional materials. For instance, holograms of black holes have been reported in graphene flakes \cite{Chen2018PRL}, ``hairy'' versions of the BTZ black hole have been suggested to exist in curved graphene sheets \cite{Kandemir2020AP}, and even wormholes seem to exist between neighboring sheets of graphene \cite{Garcia2020NPB}. Due to the attractive and promising applications of 2-dimensional materials in nanotechnology, it is thus not a surprise that engines fueled with such materials are a subject of intense study \cite{Pena2015PRE,Pena2020PRE,Singh2021PRB}. Our present results contribute to this vivid and exciting field of research, and they may pave the way for more investigations of thermodynamic consequences of curved space-time.

\acknowledgements{E.F. gratefully acknowledges support from UMBC through an Undergraduate Research Award. S.D. acknowledges support from the John Templeton Foundation under Grant No. 62422.}

\bibliographystyle{eplbib}
\bibliography{BH_engine}

\end{document}